\documentclass[twocolumn,letterpaper,10pt]{article}
\usepackage{iasted}
\usepackage{times}
\usepackage[dvips]{graphicx}

\title{A new authentication protocol for revocable anonymity in ad-hoc networks}

\author{Adam Wierzbicki \\ Polish-Japanese Institute \\of Information Technology\\Warsaw, Poland \\adamw@pjwstk.edu.pl
\and Aneta Zwierko$^1$ \\
$^1$Warsaw University \\of Technology, 
Institute \\of Telecommunication\\
Warsaw, Poland\\
azwierko@tele.pw.edu.pl
\and Zbigniew Kotulski$^{1,2}$ \\ 
$^2$Polish Academy of Sciences \\
Institute of Fundamental \\
Technological Research\\
Warsaw, Poland\\
zkotulsk@ippt.gov.pl}

\date{}
\begin{document}

\maketitle
\thispagestyle{empty}

\begin{abstract}
This paper describes a new protocol for authentication in ad-hoc networks. The protocol has been designed to meet specialized requirements of ad-hoc networks, such as lack of direct communication between nodes or requirements for revocable anonymity. At the same time, a ad-hoc authentication protocol must be resistant to spoofing, eavesdropping and playback, and man-in-the-middle attacks. The article analyzes existing authentication methods based on the Public Key Infrastructure, and finds that they have several drawbacks in ad-hoc networks. Therefore, a new authentication protocol, basing on established cryptographic primitives (Merkle's puzzles and zero-knowledge proofs) is proposed. The protocol is studied for a model ad-hoc chat application that provides private conversations.
\end{abstract}
\section{Introduction}

Authentication services are required by many applications of ad hoc networks, both mobile (MANETs) or wired, like peer-to-peer. As an example, consider chats, games, or data sharing in a ad-hoc network, or in a MANET. As more practical applications of MANETS will be developed, the need for authentication services will grow. In addition, many forms of secure routing in MANTETs or general ad-hoc networks cannot operate without a form of authentication.

At the same time, ad-hoc networks and their applications are more vulnerable to a number of well-known threats, such as identity theft (spoofing), violation of privacy, and the man-in-the-middle attack. All these threats are difficult to counter in an environment where membership and network structure are dynamic and the presence of central directories cannot be assumed. 

Applications of ad-hoc networks can have ano\-ny\-mi\-ty requirements that cannot be easily reconciled with some forms of authentication known today. On the other hand, service providers that are  bound by legal regulations have to be able to trace the actions of user of a MANET. Finding a reasonable trade-off between these two requirements is rather hard. In this paper, we use the term \emph{revocable anonymity} for a system in which a user cannot be identified to the outside world, but a trusted authority is provided with the possibility to identity actions performed by each user.

These considerations lead to the conclusion that mobile ad hoc networks can benefit from new, specialized methods of authentication. In this article, we combine two cryptographic techniques - Merkle's puzzles and zero-knowledge proofs - to develop a protocol for authentication in ad-hoc networks. This protocol is resistant to man-in-the-middle and eavesdropping attacks and prevents identity theft. However, the protocol allows for revocable anonymity of users and is adapted to the dynamic and decentralized nature of these networks. Finally, our protocol works with any MANET routing protocol and does not assume any properties of MANET routing.

We study the protocol for a model chat application in an ad-hoc network that needs to authenticate users to continue concurrent private conversations. Users of the chat prefer to remain anonymous, but they must have identities for the duration of the conversation. However, the applications of an authentication protocol in ad-hoc networks can be wide, and our authentication protocol can be adapted to many other applications. In this paper, we aim to demonstrate the principle that lies behind the new authentication method, to compare the new method to existing techniques and to analyze its security and performance.

\paragraph{Organization of paper.} In the next section, we consider how existing techniques such as Public Key Infrastructure or their modifications can be used for authentication in MANET applications. We present a simple case study of a chat application. We demonstrate how the use of PKI is difficult if users have no previous knowledge of the receiver of messages. Other disadvantages of PKI are the lack of global availability and the lack of anonymity. In section~\ref{crypto}, we present and explain the cryptographic primitives used in our protocol. Section~\ref{prot} presents the protocol, and concludes with an analysis of the protocol's security and efficiency. Section~\ref{concl} concludes and discusses further work.

\section{Related work}

General security architectures for MANETs almost exclusively use public key cryptography (PKI or the Web of trust) \cite{confidant, adhoc-PGP, adhoc-book}. These systems provide authentication without anonymity, and will be discussed in more details below.

Most systems that provide anonymity are not interested in allowing to trace the user under any circumstances. \emph{Chaum mixing networks}, proxy servers, have not been designed to provide accountability. For mobile ad hoc networks, approaches exists that provide unconditional anonymity, again without any accountability \cite{adhoc-onion}.

A Chaum mixing network, mentioned earlier, is a collection of special hosts (mixing nodes) that route user messages. Each node simply forwards an incoming messages to other nodes in the mixing network. The path (sequence of nodes) is chosen by the sender, and the message is put into envelopes (based on PKI infrastructure), one for each node on the path.

An area that requires both anonymity and accountability is agent systems (\cite{NIST}). Most of the security architectures for those systems do not provide any anonymity, e.g.,~\cite{C},~\cite{G},~\cite{F}. 

However there exists work devoted to different anonymity aspects e.g.~\cite{KK2}.

A different scheme that preserves anonymity is proposed in~\cite{D}. The scheme is based on a credential system and offers an optional anonymity revocation. Its main idea is based on oblivious protocols, encryption circuits and the RSA assumption. 

\section{Authentication in MANET applications}

In this section, we discuss a model chat application in a MANET that will be used to guide our discussion of authentication. However, the conclusions of the discussion apply to any MANET application that requires authentication.

\subsection{Chat of users in a MANET}
Consider a chat application in a mobile ad-hoc network. The system makes it possible for users to execute two operations: $SALL(m)$ and $SPRIV(u, m)$. The first operation sends a message, $m$ to all users in the network. The second operation sends a private message $m$ to a selected user, $u$. Note that a user that executes the $SALL$ operation needs not to know who receives the message.

\begin{figure}[h]
\centering
\includegraphics[width=8cm]{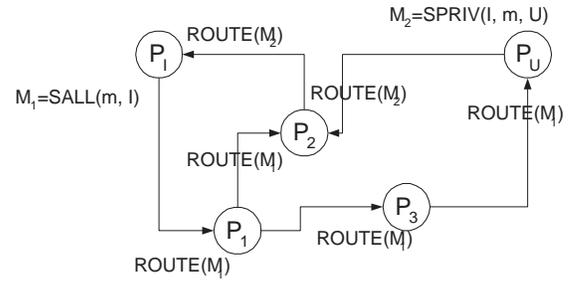}
\caption{Chat of users in a MANET}
\label{fig1}
\end{figure}

The described system is visualized on figure~\ref{fig1}. The $SALL$ and $SPRIV$ messages are routed by the network using the $ROUTE$ operation (using any MANET routing protocol). On the figure, the $SALL$ message is routed from user $P_I$ to all other users, among them to $P_U$ by the nodes $P_1$ and $P_3$. 
After that, $P_U$ responds by sending a message $SPRIV$ to $P_I$. The exchange of private messages may continue concurrently to the sending of messages to all users by either $P_I$ or $P_U$.

Consider now that the application wishes to authenticate the private message senders. For example, after receiving the first $SPRIV$ message, the application creates a file that will contain all messages exchanged by the two users during a private conversation. Only the user that has sent the first $SPRIV$ message is authorized to continue the conversation. In order to enforce this, some form of authentication is required.

The first question is whether the user address in the MANET is sufficient for authentication. Is it possible for a malicious user, $P_M$, to assume the address of an innocent user, $P_I$? In MANETs, the possibility of successfully spoofing an IP address cannot be overlooked. 

Let us assume that $P_I$ uses his own address as authentication information. The $SPRIV$ message takes the form of $SPRIV(receiver_address, m, sender_address)$. However, $P_M$ runs a DoS attack against $P_I$, forcing $P_I$ to leave the network. After $P_I$ has left, $P_M$ joins the network assuming the address of $P_I$. Next, $P_M$ can take over the private conversation of $P_I$. 

What is needed to implement access permissions that allow private conversations? An authentication mechanism that 
\begin{itemize}
	\item allows a user to authenticate its conversation partners
	\item does not use centralized control during authentication
	\item is safe against playback attack
	\item is safe against eavesdropping
	\item is safe against man-in-the-middle attack
	\item provides controlled anonymity
\end{itemize}

\subsection{Case Study: $PKI$}

Before we present a new method of authentication, let us first describe and analyze available means of providing authentication in MANETs. The most well known (and most frequently used) method is authentication using public key cryptography and Public Key Infrastructure ($PKI$) certificates. If such a method is used in a MANET, all users must obtain a $PKI$ certificate from a certificate authority ($CA$) in order to access certain system functions (perhaps some functions may be available without access control).

An alternative would be to use a trusted source of authentication information that is part of the MANET: a bootstrap server. This element (we shall refer to it as authenticating bootstrap, $AB$) issues certificates to users that join the system. A drawback of this approach is that the identity of users is not externally verified. A similar approach is to allow all users to issue certificates like in the PGP "Web of trust" model. In \cite{ESpeak}, this approach has been chosen along with the use of SPKI; however, this work does not significantly differ from the approach described in this case study.

\begin{figure}[h]
\centering
\includegraphics[width=8cm]{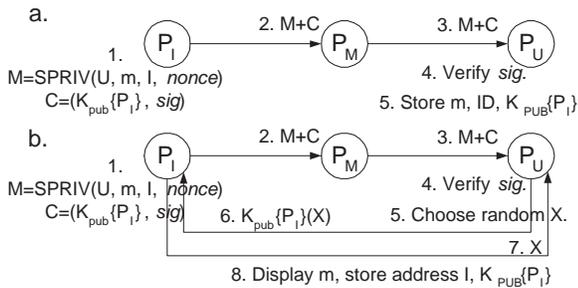}
\caption{Using certificates for authentication in a MANET application}
\label{fig2}
\end{figure}

Let us consider, how the described authentication methods could be used to solve the problem posed above: implementing access permissions for a chat with private conversations in a MANET. A proposed solution is shown on fig.~\ref{fig2}. 

$P_I$ has a certificate, $C$, that contains its public key and a signature, $SIG$. The certificate and the $SPRIV$ message are routed through the MANET (the message contains a nonce to avoid playback attacks). For simplicity, let us assume that there is a single, malicious user on the path from $P_I$ to $P_U$. When $P_U$ receives the message, he can verify the validity of the signature and accept the public key of $P_I$ as authentication information. In the future, $P_U$ will only display private messages from $P_I$ if the message has been signed by $P_I$. Verification of the certificate may require communication with $CA$ or $AB$, if $P_U$ does not know the public key of the $CA$ or $AB$.

However, note that the presented scenario is insecure. $P_M$ is capable of a man-in-the-middle attack that exchanges the certificate, $C$, with a certificate of $P_M$, $C'$, and the address of $P_I$ with the address of $P_M$. Unless the receiver, $P_U$, is capable of verifying that the certificate belongs to the sender $P_I$, then $P_M$ will be able to continue the conversation of $P_I$ afterwards (and $P_I$ will not!). To fix this problem, the proposed protocol has to be modified as presented on fig.~\ref{fig2}. The only way for $P_U$ to make sure that the certificate belongs to $P_I$ is to communicate with $P_I$ over a channel that is not controlled by $P_M$ and receive a proof that $P_I$ has a private key that matches the public key in the certificate.

After the private conversation has been accepted by $P_U$, $P_I$ may wish to send messages using another $SPRIV$ operation. For the second time, authentication can be simpler. $P_I$ and $P_U$ now know the public keys of each other. This information, or a secret value associated with the conversation during initiation, is enough to establish an encrypted channel between $P_I$ and $P_U$ and to authenticate $P_I$.

\subsection{Disadvantages of using $PKI$ for authentication in MANETs}

However, the proposed solution has several drawbacks:
\begin{enumerate}
	\item It requires direct communication with $P_I$. This may not be possible if $P_I$ is outside the radio range of $P_U$.
	\item The certificate must contain the address of $P_I$ (or the system must include a directory where this address may be found). This requires updates whenever $P_I$ changes its address. 
	\item Communication with the $CA$ or $AB$ must occur during every transaction, if $P_U$ does know the public key of $CA$ or $AB$.
	\item It requires a 3-way exchange of information.
	\item If $PKI$ certificates are used, the users cannot be anonymous.
	\item Note that we do not consider how to provide message integrity during communication from $P_I$ to $P_U$. We focus solely on authentication.
\end{enumerate}

As pointed out in~\cite{DIN}, the use of $PKI$ for authentication has other drawbacks. The use of $PKI$ is difficult because of the necessity of verifying legal identities of all participants. This is a difficult task, and may limit the participation of users from countries or geographical areas where the access to $PKI$ infrastructure is limited. Other users may have privacy concerns, depending on the type of application.
 
Finally, the security of $PKI$ has been questioned due to its hierarchical nature. In~\cite{BUR}, authors observe that if a high-level certification authority is compromised, then the result is a failure of a large part of the system. For these reasons, it may be worthwhile to consider a more lightweight, scalable and robust authentication mechanism for MANET applications.

\section{Proposal}

In this paper, we describe a new protocol for authentication for MANET applications. The protocol allows users to securely send private messages to another user (as described in section $3$).

First, utilized cryptographic primitives are briefly introduced: the concept of zero-knowledge proofs and Merkle's puzzles. Then, we present the authentication protocol.

\subsection{Cryptographic primitives}
\label{crypto}

Our scheme involves two cryptographic primitives: Merkle's puzzles and zero-know\-led\-ge proofs. We describe them shortly below.

\paragraph{Merkle's puzzles} Ralph Merkle introduced his concept of cryptographic puzzles in~\cite{M}. The goal of this method was to enable secure communication between two parties: A and B, over an insecure channel. The assumptions were that the communication channel can be eavesdropped (by any third party, called E). Assume that A selected an encryption function ($F$). $F$ is kept by A in secret. A and B agree on a second encryption function, called $G$:
\begin{center}
\emph{G(plaintext, some key) = some encrypted message}.
\end{center}

$G$ is publicly known. A will now create $M$ puzzles (denoted as $s_i$, $0 \leq i \leq M$) in the following fashion: 
\begin{displaymath}
s_i = G((K,X_i,F(X_i)),R_i)
\end{displaymath} 
$K$ is simply a publicly known constant term, which remains the same for all messages. The $X_i$ are selected by A at random. The $R_i$ are the "puzzle" part, and are also selected at random from the range $(M \cdot (i-1), M \cdot i)$. B must guess $R_i$. For each message, there are $N$ possible values of $R_i$. If B tries all of them, he is bound to chance upon the right key. This will allow B to recover the message within the puzzle: the triple $(K,X_i,F(X_i))$. B will know that he has correctly decoded the message because the constant part, K, provides enough redundancy to insure that all messages are not equally likely. Without this provision, B would have no way of knowing which decoded version was correct, for they would all be random bit strings. Once B has decoded the puzzle, he can transmit $X_i$ in the clear. $F(X_i)$ can then be used as the encryption key in further communications. B knows $F(X_i)$ because it is in the message. A knows $F(X_i)$ because A knows $X_i$, which B transmitted in the clear, and also knows F, and so can compute $F(X_i)$. E cannot determine $F(X_i)$ because E does not know F, and so the value of $X_i$ tells E nothing. E's only recourse is to solve all the $N$ puzzles until he encounters the 1 puzzle that B solved. So for B it easy to solve one chosen puzzle, but for E is computationally hard to solve all $N$ puzzles.
\label{merkle}

\paragraph{Zero-knowledge proofs}

A zero knowledge proof system (\cite{P}, \cite{ID}, \cite{FS}, \cite{OG2}, \cite{OG4}, \cite{BDLP}) is a protocol that  enables one party to \emph{prove} the possession or knowledge of a "secret" to another party, without revealing anything about the secret, in the information theoretical sense. These protocols are also known as minimum disclosure proofs. Zero knowledge proofs involve two parties: the prover who possesses a secret and wishes to convince the verifier, that he indeed has the secret. As mentioned before, the proof is conducted via an interaction between the parties. At the end of the protocol the verifier should be convinced only if the prover knows the secret. If, however, the prover does not know it, the verifier will be sure of it with an overwhelming probability.  

The zero-knowledge proof systems are ideal for constructing identification schemes. A direct use of a zero-knowledge proof system allows unilateral authentication of P (Peggy) by V (Victor) and require a large number of iterations, so that verifier knows with an initially assumed probability that prover knows the secret (or has the claimed identity). This can be translated into the requirement that the probability of false acceptance be $2^{-t}$ where $t$ is the number of iterations. A zero knowledge identification protocol reveals no information about the secret held by the prover under some reasonable computational assumptions.

\subsection{The authentication protocol}
\label{prot}

The proposed protocol offers an authentication method for the model MANET chat application. The node that wishes to send a private message is equipped with a zero-knowledge value. After the setup of a private conversation, this value will enable only the right node to send new private messages. Using the proposed protocol, the authentication information cannot be used by a node that routes the message for its own purpose. A short overview is presented in this section and a detailed description in the next.

The proposed protocol has three phases:
\begin{enumerate}
	\item initial: when a bootstrap creates necessary values for authentication
	\item initialization of private conversation: the first private message contains additional zero-knowledge values that will enable the sender (and no one else) to continue the private conversation.
	\item exchange of private messages: the sender uses a zero-knowledge proof and Merkle's puzzles to authenticate itself and to safely send a private message.
\end{enumerate}

The node that initializes the private conversation is denoted as $P_I$, the receiving node as $P_S$ and nodes that route the message as $P_1, P_2, \ldots$, the message as $m'$ (first message) and $m'', m''', \ldots$ (next messages). $A$ is the authentication data.

In this basic scenario we assume that routing nodes do not modify the data, just forward it correctly. Attacks: scenarios where these nodes can modify or eavesdrop information are described in section~\ref{sec}.

\paragraph{Phase 1 - initial}

This proposal is not directly based on zero-knowledge protocols, but on an identification system based on a zero-knowledge proof. We choose the GQ scheme (\cite{GQ}) as the most convenient for our purposes. In this scheme, the bootstrap has a pair of RSA-like keys: a public $K_P$ and a private one $k_p$. The bootstrap also computes public modulus $N = p \cdot q$, where $p, q$ are RSA-like primes. The following equation has to be true:
\begin{displaymath}
K_P \times k_p \equiv 1 (\textrm{mod }(p-1)\cdot(q-1)).
\end{displaymath}
The pair ($K_P$, $N$) is made public. The keys can be used for different purposes, not only for our system.

The bootstrap computes a set of so-called identities, denoted by $ID$, and their equivalencies, denoted by $J$. It does not matter how $J$ is obtained if it is obvious for all participants how to obtain $J$ from $ID$. The pairs $(ID, J)$ are generated for every node that requests them. The identity is used to authenticate $P_I$ during an attempt to continue the conversation. The bootstrap also computes a secret value for each $ID$: 
\begin{displaymath}
\sigma \equiv J^{-k_p} (\textrm{mod }N).
\end{displaymath}
The secret $\sigma$ is used by $P_I$ to compute correct values for the $GQ$ authentication scheme. $P_I$ obtains the following information in the initial phase: $ID$ (public) and $\sigma$ (secret).

To preserve anonymity, node $P_I$ should request at least a few different pairs $(ID, \sigma)$ or, if possible, obtain a new pair for each private conversation (key). 

\paragraph{Phase 2 - initialization of the private conversation}

The purpose of this phase is to associate a proper $ID$ with the conversation. Different methods may be used for that purpose, depending on the security and performance requirements of the system.

Here are some possibilities:
\begin{enumerate}
	\item The node $P_I$ can simply send the $ID$ with the message $m$ in open text. 
	\begin{figure}[h]
		\centering
		\includegraphics[width=8cm]{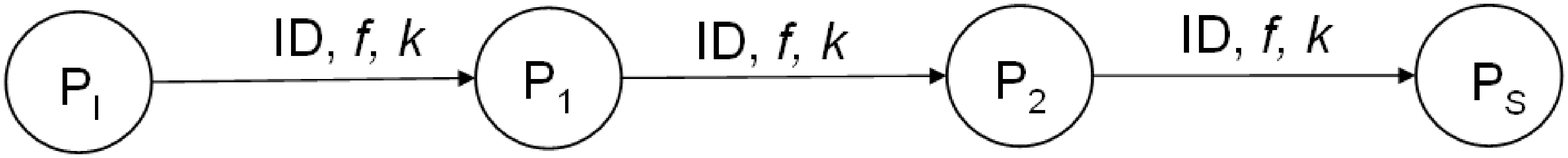}
		\label{fig3}
	\end{figure}
In that situation, the node $P_I$ has to trust all other nodes that they do not change neither message nor $ID$.
	\item The node $P_I$ can ask the bootstrap to store an $ID$ value associated with the conversation. During conversation initialization, $P_S$ contacts the bootstrap and obtains the proper $ID$. 
	\begin{figure}[h]
		\centering
		\includegraphics[width=8cm]{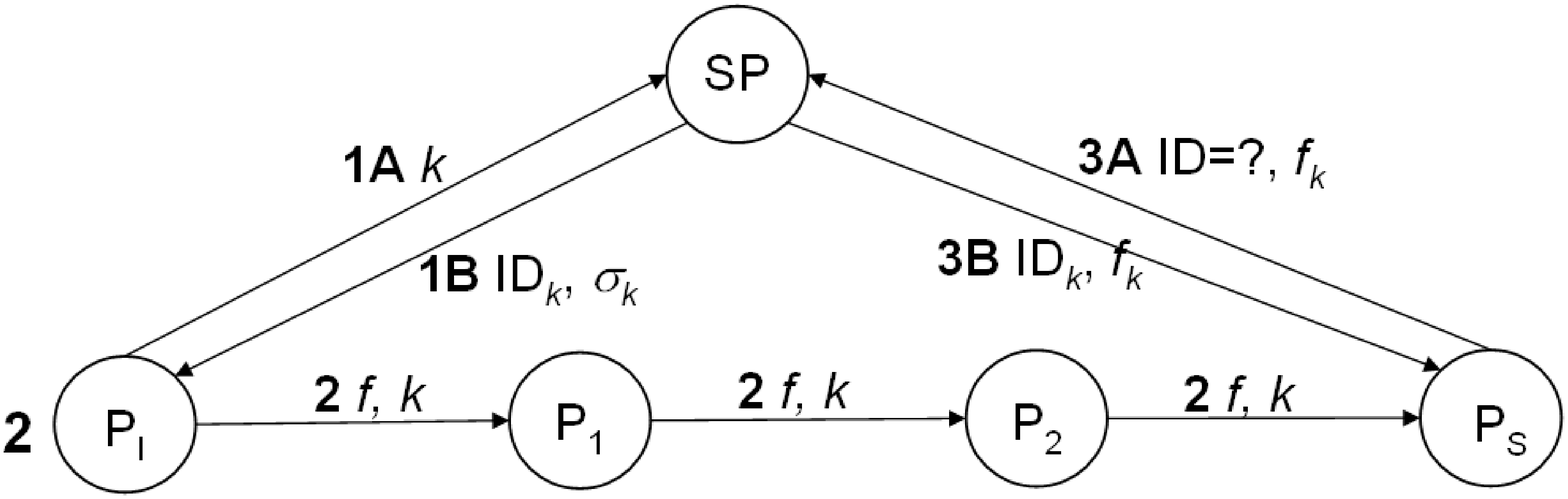}
		\label{fig4}
	\end{figure}
	 
	\item A more secure way is to use the bootstrap's keys for a different purpose, not only for the zero-knowledge protocol. After creation of an $ID$ for node $P_I$ in the initial phase, the bootstrap can sign the $ID$ with his private key. In this case, the $ID$ can be sent securely over multiple nodes. After receiving the first message, $P_S$ can check the validity of the bootstrap's signature and accept only a valid $ID$. To provide message integrity, the bootstrap would have to sign a hash of the message ($h(m)$), as well.
	\begin{figure}[h]
		\centering
		\includegraphics[width=8cm]{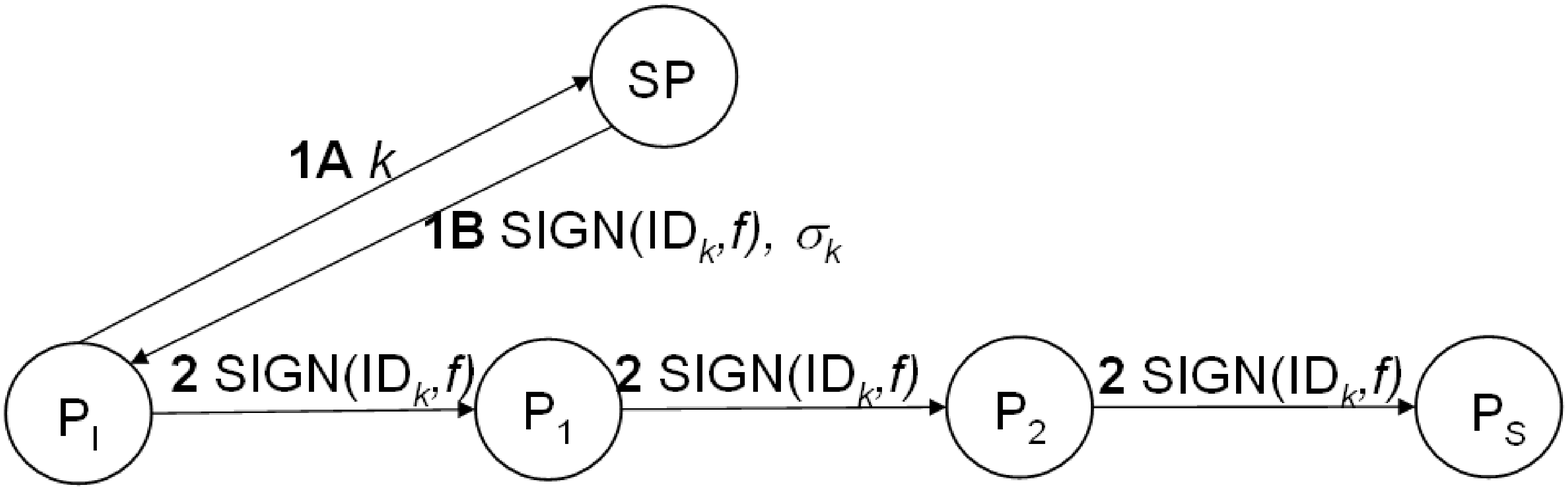}
		\label{fig5}
	\end{figure}
\end{enumerate}

\paragraph{Phase 3 - exchange of private messages}

\begin{enumerate}

	\item The node $P_I$ creates a set of puzzles: $S = \{s_1, \ldots, s_n\}$. Each puzzle has a zero-knowledge challenge. This challenge is a number computed basing on a random value $r$, $r \in \{1, \ldots, N - 1\}$. It is computed as following:
\begin{equation}
u = r^{K_P} \textrm{ (mod } N).
\label{eq1}
\end{equation}
	\textbf{Creating a set of puzzles}\\
	Each puzzle used in the proposed scheme has a following form: $G(K,X_i,F(X_i),u),R_i)$, where $K$, $X_i$, $R_i$ and $F$, are described in section~\ref{merkle}.
	\begin{table}[h]
		\caption{Possible puzzles}
		\label{tab1}	
		\centering
		\begin{tabular}{ll}
		Puzzle no & puzzle \\
		\hline
		1 ($s_1$)& $G(K,X_1,F(X_1),u),R_1)$ \\
		2 ($s_2$)& $G(K,X_2,F(X_2),u),R_2)$ \\
		 & \ldots \\
		n ($s_n$)& $G(K,X_n,F(X_n),u),R_n)$ \\
		\end{tabular}
	\end{table}
	Each puzzle can contain a different $u$ value (computed from $r$), which gives additional security.
	\item The node $P_I$ sends the whole set of puzzles to $P_S$. 
		\begin{figure}[ht]
			\centering
			\includegraphics[width=8cm]{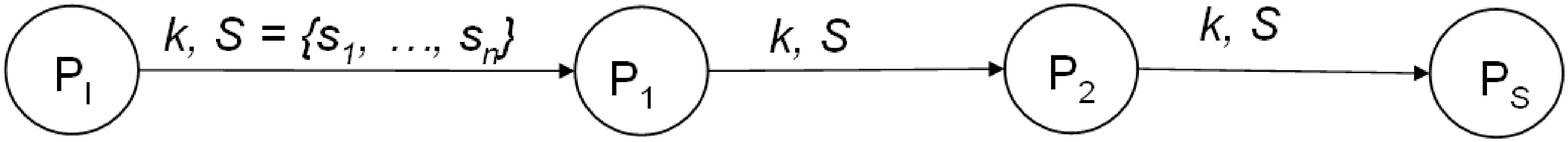}
			\label{fig6}
		\end{figure}
\newpage
	\item $P_S$ solves a chosen puzzle and chooses a random value $b \in \{1, \ldots, N\}$. $P_S$ sends the puzzle's number ($X_i$) and $b$ to $P_I$.
		\begin{figure}[h]
			\centering
			\includegraphics[width=7cm]{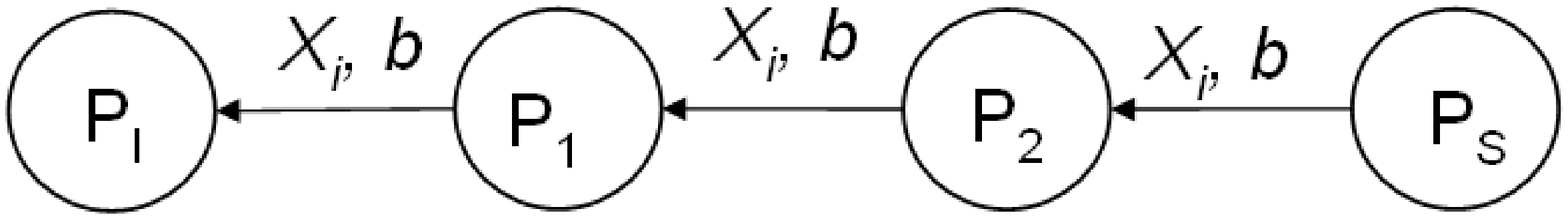}
			\label{fig7}
		\end{figure}	
	\item The $P_I$ computes the next value in the $GQ$ scheme, $v$. This values is based on the number $b$ received from $P_S$ and on the secret value $\sigma$ of $P_I$:
	\begin{equation}
	v \equiv r \times \sigma^b \textrm{ (mod } N).
	\label{eq2}
	\end{equation}
	\item $P_I$ sends $v$ and a new message (encrypted, using information from the puzzle). Some possible methods of securing the message are described below. The secured message has the form:

	\begin{displaymath}
	L(m', F(X_i)).
	\end{displaymath}
		\begin{figure}[h]
			\centering
			\includegraphics[width=8cm]{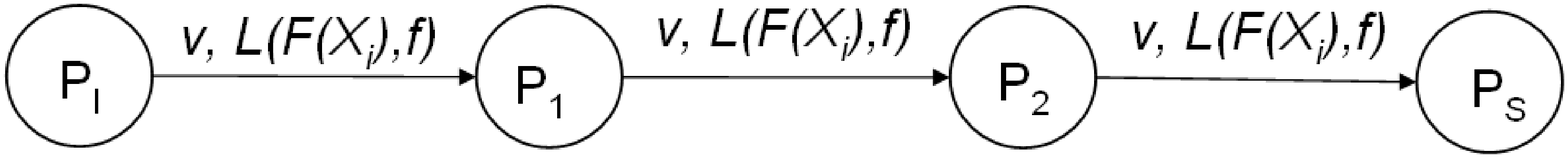}
			\label{fig8}
		\end{figure}

	\item $P_S$ uses information extracted from the puzzle, $ID$, to obtain $J$ and verify if $v$ is the right value. To validate the response from $P_I$, $P_S$ checks if
	\begin{equation}
	J^b \times v^{K_P} \equiv u \textrm{ (mod } N).
	\label{eq3}
	\end{equation}
	If the equation is satisfied, then the new message is accepted. 
\end{enumerate}

\paragraph{Securing the new message} 
The value $F(X_i)$ is a secret known only to $P_S$ and $P_I$. Thus, it can be used to establish a secure channel for the new message. This can be used to provide:
\begin{enumerate}
	\item encryption: the message could be encrypted using $F(X_i)$ as a key for a symmetric cipher.
	\item integrity: the hash of the message could be encrypted using a symmetric cipher with key $F(X_i)$. 
\end{enumerate}
\label{secfile}

\subsection{Security of proposed scheme}
\label{sec}

In this section, we are going to discuss only the security of phase 3 of the protocol, since the protocol offers several distinct possibilities in phase 2, each one with a different level of security.

\paragraph{Continuing the conversation by an unauthorized user}

Assume that one of the routing nodes ($P_M$) wishes to send a private message and impersonate $P_I$. $P_M$ cannot impersonate $P_I$, because $P_M$ does not have the $\sigma$ value to obtain the correct $v$ for the authentication phase of the protocol. The values $u$, $b$ or $ID$ do not contain any information that would be useful in cheating $P_S$. This property is assured by the zero-knowledge protocol.

The message itself is secured by methods described in sec~\ref{secfile}, using the $F(X_i)$ value. For any eavesdropper it is computationally infeasible to solve all puzzles to find the puzzle the with proper $X_i$ (the one used by $P_S$) if the number of puzzles is large enough. E.g. if the function $G$ would be $DES$, and $R_i$ would be a key for a cipher with fixed 24 bits (so efficiently 32 bits long), then the number of computations required to solve one puzzle is about $2^{31}$. Now it is easy to estimate how many puzzles should be created by $P_I$.

\paragraph{Eavesdropping}

An eavesdropping node, $P_M$, can observe all values of the zero-knowledge protocol: $u$, $b$ and $r$. This knowledge does not reveal anything about the secret $\sigma$ and since $u$ and $b$ are random and change in every iteration of the protocol, that does not enable $P_M$ to interfere and gain any important information. Also, if the number of puzzles is sufficiently large, solving all puzzles is infeasible in reasonable time and finding the puzzle that was used to secure the message is hard.

\paragraph{Play-back attack}

Using our protocol, $P_S$ chooses a random value $b$ and then $P_I$ has to compute the $v$ value, which is later utilized by $P_S$ to check if the authentication is successful. Therefore, previously used $v$, $r$ ($u$) and $b$ values are useless. Only $P_I$ is able to create the proper $v$ value for a random $b$.

\paragraph{Man-in-the-middle attack} The goal of this attack is to either change the new message or to gain some information about $\sigma$ by one of the intermediate nodes ($P_M$). A property of zero-knowledge proofs used in our protocol is that gaining any information about $u$, $b$ and $v$ values does not reveal anything about the $\sigma$. Changing the message is also not possible since it is protected with the secret value $F(X_i)$, known only to $P_I$ and $P_S$. 

\subsection{Performance analysis}

In the proposed system there are two phases when computational overhead could be significant:
\begin{enumerate}
	\item computing values for the zero-knowledge protocol (equations:~\ref{eq1},~\ref{eq2},~\ref{eq3}). The number of computations needed for these equations is similar to computations of public key cryptography. The $u$ value (eq.~\ref{eq1}) can be calculated offline.
	\item computing the set of puzzles: this also can be done by $P_I$ offline. We assume that the $G$ function would be $DES$ or any other symmetric cipher, so it would be quite fast to compute a single puzzle. The amount of computations depends rather on the number of puzzles ($M$) and is similar to encrypting a message of size $M \cdot n$, where $n$ is the size of $N$ in bits (because $u < N$). 
	\item Sending the set of puzzles is the only significant communication overhead. The size of the set of puzzles depends on the required security level and is difficult to estimate (without additional assumptions about available computational power of malicious nodes). Moreover, since the breaking of all puzzles should take more time than the transmission of the entire message, perhaps the size of the set of puzzles could depend on the message size (be bounded above by a fraction of the message size, for instance $1\%$).

\end{enumerate}

\subsection{Comparison with PKI}

Let us compare our protocol using the same criticism as for \emph{PKI} in section 4:

\begin{enumerate}
	\item direct communication of $P_I$ and $P_S$ is no longer required
	\item a directory or method to obtain the address of $P_I$ by $P_S$ is not necessary
	\item communication with the bootstrap may be required during the initialization of a private conversation, depending on the chosen method of communicating the $ID$
	\item 3-way exchange of information is not required during conversation initialization, but only for subsequent messages
	\item the proposed protocol provides revocable a\-no\-ny\-mi\-ty.
\end{enumerate}

\section{Conclusions}
\label{concl}

Authentication in P2P/ad-hoc systems is surprisingly difficult due to the fact that nodes often do not know the identity of each other before they communicate. In a client-server system, at least the identity of the server is known to the client. This simplifies the use of \emph{PKI} for authentication. In a P2P/ad-hoc system, the use of \emph{PKI} may require direct communication of two nodes to prevent a man-in-the-middle attack.  This is difficult to realize in a P2P/ad-hoc system.

We have developed an authentication method that is secured against eavesdropping, man-in-the-middle, and playback attacks in a P2P/ad-hoc system, but does not require direct communication. The proposed method does not introduce significant computational or communication overheads. Also, the proposed method provides revocable anonymity that is not available when \emph{PKI} is used.

The proposed system of authentication with revocable anonymity gives quite new possibilities for security solutions in P2P/ad-hoc networks. First, it provides anonymity of the operating node against other nodes and any external users, except for the bootstrap. Additionally, the system makes it possible to identify a node's actions when cooperating with the bootstrap. If practically implemented, the system can be controlled against a malicious nodes trying violate the rules of a MANET application. The applications of this control can range from games in a MANET to prevention of indecent or malicious messages on MANET chats.

\paragraph{Future work}
The form of anonymous authentication and revocable anonymous authentication should perhaps depend on the particular MANET application. Thus, the first possible extension of the results presented here is a precise analysis of requirements of chosen MANET applications. This problem will be the subject of future research.  

Another extension of the presented results is to offer new security services. The first natural proposition is mutual authentication of nodes, then non-repudiation of operations and finally combinations of all common security services applied to nodes and the routed messages.

\end{document}